\documentclass[twocolumn,showpacs,preprintnumbers,amsmath,amssymb]{revtex4}
\usepackage{graphicx}
\usepackage{dcolumn}
\usepackage{bm}
\begin{document}
\title {Asymptotic near nucleus structure 
of the electron-interaction potential
in local effective potential theories}
\author {Zhixin Qian}
\affiliation{Department of Physics,
Peking University, Beijing 100871, China}
\author{Viraht Sahni}
\affiliation{Department of Physics, Brooklyn College and
the Graduate School, City University 
of New York, Brooklyn, NY 11210, USA}
\date{\today}
\begin{abstract}
In local effective potential theories of 
electronic structure, the electron correlations
due to the Pauli exclusion principle, Coulomb repulsion, and 
correlation-kinetic effects, are all incorporated in the local
electron-interaction potential $v_{ee}({\bf r})$. In previous work, it
has been shown that for spherically symmetric or sphericalized
systems, the asymptotic near nucleus expansion of this potential
is $v_{ee}(r) = v_{ee}(0) + \beta r + O(r^2)$, 
with $v_{ee}(0)$ being finite. By assuming that the
Schr\"odinger and local effective potential
theory wave functions are analytic near the nucleus of atoms,
we prove
the following via Quantal density functional theory (Q-DFT): (i)
correlations due to the Pauli principle and Coulomb correlations
do not contribute to the linear structure; (ii) these Pauli and Coulomb
correlations contribute quadratically; (iii) the linear structure is 
{\em solely} due to correlation-kinetic effects, 
the contributions of these
effects being determined analytically. We also derive
by application of adiabatic coupling constant perturbation
theory via Q-DFT (iv) the asymptotic near nucleus expansion 
of the Hohenberg-Kohn-Sham theory exchange $v_x({\bf r})$ and correlation
$v_c({\bf r})$ potentials. These functions also approach the nucleus
linearly with the linear 
term of $v_x({\bf r})$ being {\em solely} due to the
lowest-order correlation kinetic effects, and the linear term of 
$v_c({\bf r})$ being due {\em solely} to the higher-order correlation
kinetic contributions. The above conclusions are equally valid for
systems of arbitrary symmetry, provided spherical averages of the
properties are employed.
\end{abstract}
\maketitle

\section{Introduction}
The basic idea underlying all 
time-independent 
{\em local} effective
potential theories of electronic structure such as 
Hohenberg-Kohn-Sham density functional theory (HKS-DFT) 
\cite{HKS} and Quantal density functional theory (Q-DFT) \cite{sahni},
is the mapping from a system of $N$ electrons in some external field
${\bf {\cal F}}^{ext}({\bf r}) = - \bigtriangledown v({\bf r})$,
where $v({\bf r})$ is the external potential, to one
of noninteracting Fermions with equivalent density $\rho({\bf r})$.
For example, the
potential energy of the electrons in the external field of
the nucleus of atoms is $v({\bf r})= -Z/r$, where $Z$ is the atomic
number. The Schr\"odinger equation for the electrons is (in a.u.)
\begin{eqnarray} \label{schrodinger}
{\hat H} \Psi = ({\hat T} + {\hat V} + {\hat U})\Psi = E \Psi ,
\end{eqnarray}
where ${\hat T}= - \frac{1}{2} \sum_i \bigtriangledown_i^2$,
${\hat V}= \sum_i v({\bf r}_i)$, ${\hat U}= \frac{1}{2} \sum_{i \neq j}
\frac{1}{|{\bf r}_i - {\bf r}_j|}$, $\Psi$ the wavefunction, and $E$
the energy. The density of the electrons is $\rho({\bf r}) = \langle \Psi
| {\hat \rho} | \Psi \rangle$, where
the Hermitian operator ${\hat \rho}= \sum_i \delta({\bf r}_i - {\bf r})$.
The corresponding
Schr\"odinger equation for the model electrons ({\em S} system) is
\begin{eqnarray} \label{KSequation}
[-\frac{1}{2} \bigtriangledown^2 
+ v({\bf r}) +v_{ee}({\bf r})] \phi_i({\bf r})
= \epsilon_i \phi_i({\bf r}); i = 1, \dots, N .
\end{eqnarray} 
The {\em S} system wavefunction is the Slater 
determinant $\Phi\{\phi_i \}$, so that
$\rho({\bf r}) = \langle \Phi | {\hat \rho} | \Phi \rangle =
\sum_i \phi_i^*({\bf r})\phi_i({\bf r})$. The $v_{ee}({\bf r})$ is the local
electron-interaction potential in which all the many-body
correlations are incorporated. 
The correlations are those due to the Pauli
exclusion principle and Coulomb repulsion, and the correlation-kinetic effects
that represent the difference in kinetic energy betweeen the interacting
and noninteracting systems. The highest occupied eigenvalue $\epsilon_m$
of Eq. (\ref{KSequation}) is the negative of the removal energy
whether it is the ionization potential of atoms and molecules or the 
work function of metals \cite{sahni,perdew,levy}.

It is evident that any rigorous knowledge of the exact structure of the 
potential $v_{ee}({\bf r})$, and an understanding of how the
different electron correlations contribute to this structure, is
critical to the accurate construction of the model system. 
In our previous work \cite{qs} we had shown that $v_{ee}(r)$
is finite at the nucleus for spherically symmetric system.
(For more general case of arbitrary symmetry, see Ref. \cite{pan}.) 
We assume that the asymptotic
near nucleus structure of $v_{ee} ({\bf r})$ for spherically
symmetric systems is of the general form
\begin{eqnarray}   \label{vee}
v_{ee}(r) = v_{ee}(0) + \beta r + \gamma r^2.
\end{eqnarray}
In Eq. (\ref{vee}) and the following, ${\bf r}$ is defined as the electronic
coordinate relative to one of the nuclei in the system. $v_{ee}(0)$ represents
the value at electron-nucleus coalescence.
In this paper we prove the following via Q-DFT: (i) There are
no contributions to the linear term due to correlations arising
from the Pauli exclusion principle and Coulomb repulsion;
(ii) These Pauli and Coulomb correlations contribute to the quadratic
term; (iii) The linear term arises {\em solely} 
due to correlation-kinetic
effects, with the coefficient $\beta$ being
\begin{eqnarray}  \label{beta}
\beta=\frac{4Z}{3\rho(0)}[t(0) -t_s(0)],
\end{eqnarray}
where $t({\bf r})$ and $t_s({\bf r})$
are the kinetic energy
densities of the interacting and noninteracting $S$ systems, respectively;
(iv) Finally, employing adiabatic coupling constant perturbation theory
\cite{harris,levyqian}, we derive via Q-DFT the asymptotic near 
nucleus structure of the HKS-DFT exchange $v_x({\bf r})$ and correlation
$v_c({\bf r})$ potentials. Although the above results are derived
specifically for spherically
symmetric systems, they are also valid for arbitrary symmetry
provided a spherical average of these properties are taken. (The
results of this paper have also been obtained within the framework
of HKS-DFT \cite{qian}.)

To put the present work in context, we note that there had been a
long standing controversy about the structure of $v_{ee} ({\bf r})$
at and near the nucleus of atoms. The traditional way of determining the 
{\em exact} structure of $v_{ee} ({\bf r})$ employs methods that
assume knowledge of the true density $\rho({\bf r})$. Then one determines, 
via numerical methods, the $v_{ee}({\bf r})$ that generates the orbitals leading 
to this density \cite{Peirs}. 
The densities employed are those obtained from
correlated or configuration-interaction type wavefunctions.
While some work \cite{smith,davidson,umrigar} on the
helium atom showed $v_{ee} ({\bf r})$ to be finite
at the nucleus, work by others \cite{almbladh-pedroza,pedroza} showed it to 
be singular there. This singularity is introduced by the inaccuracies
in the numerical calculations, for instance, those in the density arising
from the limitations of basis sets employed to solve the
differential equations for the wave functions or
one-particle orbitals \cite{basis}. Another key reason \cite{pan}
for the singularity of $v_{ee}({\bf r})$ at the nucleus is the lack
of satisfaction of the electron-nucleus coalescence condition by the approximate
wave function from which the density is obtained. For the determination
of its structure for few electron molecular systems \cite{leeuwen}
such as $H_2$ and $LiH$,
the $v_{ee} ({\bf r})$ is assumed to be finite at each nucleus.
Additional
work on light atoms, showed it to 
be either finite \cite{zhao1,zhao2,filippi} or to
diverge at the nucleus \cite{chen}.
Expressions for $v_{ee}({\bf r})$ at a nucleus have also
been derived, but once again
they are based on the assumption that \cite{liu1}
it is finite there. In various approximations
within HKS-DFT \cite{singh,langreth,perdew1,perdew2,becke}, 
the $v_{ee} ({\bf r})$ also diverges at the nucleus \cite{chen}.
With the knowledge that $v_{ee}({\bf r})$ is finite at the nucleus, what
remained was to understand whether the structure of $v_{ee}({\bf r})$
approaches the nucleus linearly or
quadratically \cite{Peirs,smith,davidson,umrigar,nagysen}. 
It is the answer to this question that we provide
in this work.

We begin in Sec. II by giving the key equations of time-independent
Q-DFT. In Sec. III we provide the asymptotic near nucleus expansion
of the interacting and noninteracting system wavefunctions, 
and the corresponding expressions for the equivalent densities. The proof
that Pauli and Coulomb correlations contribute quadratically and
not linearly to the near nucleus expansion of $v_{ee}({\bf r})$ is
given in Sec. IV. In Sec. V we prove that the contribution to
the linear term of $v_{ee}({\bf r})$ is solely a consequence of
correlation-kinetic effects, and derive an expression for this
contribution. Combining the results of the previous two
sections, we present in Sec. VI the expression for the
near nucleus structure of $v_{ee}({\bf r})$. Finally, by 
employing coupling constant perturbation theory
within the framework of Q-DFT, we derive in Sec. VII the
near nucleus structure of the HKS-DFT exchange and
correlation potentials. Concluding remarks are made in Sec. VIII.

\section{Brief introduction to Q-DFT}

In time-independent Q-DFT \cite{sahni,holas}, the potential
$v_{ee} ({\bf r})$
is the work
done to move a model fermion in the force of a conservative field
${\bf \cal F}^{eff}({\bf r})$:
\begin{eqnarray} \label{vee-eee-ztc}
v_{ee} ({\bf r}) = - \int_\infty^{\bf r} {\bf \cal F}^{eff}({\bf r}')
\cdot d {\bf l}' ,
\end{eqnarray}
where ${\bf \cal F}^{eff}({\bf r})= {\bf \cal E}_{ee}({\bf r})
+{\bf \cal Z}_{t_c} ({\bf r})$.
Here ${\bf \cal E}_{ee}({\bf r})$ is the electron-interaction field 
representative of the Pauli and Coulomb correlations, and 
${\bf \cal Z}_{t_c} ({\bf r})$ the correlation-kinetic field 
representative of those
effects. 
For spherically symmetric systems, 
the fields ${\bf \cal E}_{ee}({\bf r})$
and ${\bf \cal Z}_{t_c} ({\bf r})$ are each 
conservative. Thus, $v_{ee} ({\bf r})
=W_{ee} ({\bf r}) + W_{t_c}({\bf r})$, where the work done
$W_{ee} ({\bf r})$, $W_{t_c}({\bf r})$ are separately
path-independent. 
The field ${\bf \cal E}_{ee}({\bf r})$
is obtained via Coulomb's law from its quantal source, the
pair-correlation density $g({\bf r}, {\bf r}')$, as 
\begin{eqnarray}  \label{eee1}
{\bf \cal E}_{ee}({\bf r}) = \int d {\bf r}' g({\bf r}, {\bf r}')
\frac{{\bf r} - {\bf r}'}{|{\bf r} - {\bf r}'|^3} ,
\end{eqnarray}
where 
$g({\bf r}, {\bf r}') = \langle \Psi|\sum_{i \neq j}\delta({\bf r}_i
-{\bf r}) \delta({\bf r}_j 
-{\bf r}') | \Psi \rangle/\rho({\bf r})$.
The pair-correlation density $g({\bf r}, {\bf r}')$ may also
be written as $g({\bf r}, {\bf r}')= \rho({\bf r}') +
\rho_{xc}({\bf r}, {\bf r}')$, where $\rho_{xc}({\bf r}, {\bf r}')$,
its nonlocal component is the Fermi-Coulomb hole charge
distribution. The pair-correlation density $g_s({\bf r}, {\bf r}')$
of the $S$ system may also be expressed in terms of its local
and nonlocal components: $g_s({\bf r}, {\bf r}')= \rho({\bf r}') +
\rho_{x}({\bf r}, {\bf r}')$, where the Fermi hole charge
$\rho_{x}({\bf r}, {\bf r}')
=- |\gamma_s({\bf r}, {\bf r}')|^2/2\rho({\bf r})$,
with $\gamma_s({\bf r}, {\bf r}')
= \sum_{i \sigma} \phi_i^*({\bf r}) \phi_i({\bf r}')$
the Dirac density matrix. Defining the Coulomb hole charge
$\rho_{c}({\bf r}, {\bf r}')= g({\bf r}, {\bf r}')-
g_s({\bf r}, {\bf r}')$, the field
${\bf \cal E}_{ee}({\bf r})$ may then be written in terms of
its components as
\begin{eqnarray}
{\bf \cal E}_{ee}({\bf r})={\bf \cal E}_{H}({\bf r})
+{\bf \cal E}_{x}({\bf r}) +{\bf \cal E}_{c}({\bf r}) ,
\end{eqnarray}
where the quantal sources of the 
Hartree ${\bf \cal E}_{H}({\bf r})$, 
Pauli ${\bf \cal E}_{x}({\bf r})$, 
and Coulomb ${\bf \cal E}_{c}({\bf r})$ fields
are the density $\rho({\bf r})$, Fermi hole $\rho_{x}({\bf r}, {\bf r}')$, 
and Coulomb hole 
$\rho_{c}({\bf r}, {\bf r}')$, respectively.
It follows that $W_{ee}({\bf r}) = W_{H}({\bf r})
+ W_{x}({\bf r}) + W_{c}({\bf r})$, where the Hartree $W_{H}({\bf r})$, 
Pauli $W_{x}({\bf r})$, and
Coulomb $W_{c}({\bf r})$ potentials 
are the work done in the respective fields. The 
field ${\bf \cal Z}_{t_c} ({\bf r})$ is defined as
\begin{eqnarray}   \label{Ztc1}
{\bf \cal Z}_{t_c} ({\bf r})= \frac{1}{\rho({\bf r})}
[ {\bf z}_s({\bf r}) - {\bf z}({\bf r}) ],
\end{eqnarray}
where
${\bf z}({\bf r})$ is defined in terms of the single-particle
density matrix $\gamma({\bf r}, {\bf r}')$ as 
\begin{eqnarray}   \label{z-defin}
z_i({\bf r})
= 2 \sum_j \partial t_{ij}({\bf r})/\partial r_j,
\end{eqnarray}
with 
\begin{eqnarray}  \label{tij0}
t_{ij}({\bf r}) = \frac{1}{4} [\partial^2/\partial r_i
\partial r'_j +\partial^2/\partial r_j
\partial r'_i] \gamma({\bf r}, {\bf r}')|_{{\bf r}' = {\bf r}},
\end{eqnarray}
the kinetic energy density tensor.
The single particle density
matrix is defined as
\begin{eqnarray}  \label{gamma}
\gamma({\bf r}, {\bf r}')
= N \sum_{s} 
\int && d{\bf x}_2 \dots d{\bf x}_N 
\Psi^*({\bf r} s, {\bf x}_2, \dots, {\bf x}_N)  \nonumber \\ 
&& \Psi ({\bf r}' s, {\bf x}_2, \dots, {\bf x}_N) .
\end{eqnarray}
The ${\bf z}_s({\bf r})$
is similarly defined in terms of the Dirac density matrix
$\gamma_s({\bf r}, {\bf r}')$. 

\section{Near nucleus behavior of the wavefunctions and the density}

We first present the near nucleus behavior 
of the wavefunctions for the interacting system
and the noninteracting $S$ system, and of the density.
As is customary, and the case in prior work \cite{kato,bingel},
we assume that the wave function near the nucleus is smooth. In other words, 
that the electronic wave function of a bound state for an atom or molecule is
analytic except at the nuclear cusps. This assumption requires the stringent
condition that the Hamiltonian be analytic except where two or
more particles coincide, a condition that may not be mathematically
rigorous. However, in mathematical physics, the assumption of smoothness of the 
wave function is standard practice, and we proceed by making it. Hence,
we write the ground state many-body wave function as 
\begin{eqnarray}\label{wavefunction}
\Psi({\bf r}, {\bf X})&& =  \Psi(0, {\bf X}) + a({\bf X})r
+ b({\bf X})r^2 +\dots   \nonumber \\
&& + \sum_{m=-1}^{1}[a_{1m}({\bf X})r
+ b_{1m}({\bf X})r^2] Y_{1m}({\hat r}) + \dots  \nonumber \\
&& + \sum_{m=-2}^{2}b_{2m}({\bf X})r^2 Y_{2m}({\hat r}) + \dots,
\end{eqnarray}
for small $r$, where ${\hat r}={\bf r}/r$, and
$Y_{lm}({\hat r})$ is the usual spherical harmonics. We use ${\bf X}$
for the coordinates
$s, {\bf r}_2 s_2, \dots, {\bf r}_N s_N$.
The coefficients in the wavefunction satisfy the following 
relations:
\begin{eqnarray}\label{a-Psi(0)}
&& a({\bf X}) + Z \Psi(0, {\bf X})=0 , \nonumber \\
&& 2 b_{1m}({\bf X}) + Za_{1m}({\bf X}) =0 .
\end{eqnarray}
The above relations originate from the
cusp of the wavefunction at the nucleus,
and were derived previously in Ref. \cite{kato,bingel}.
The expansion of the wavefunction is obtained
by making use of the asymptotic expansion for small $r$
\begin{eqnarray}
\frac{1}{|{\bf r} -{\bf r}_i|} = 
\sum_{l=0}^\infty \sum_{m=-l}^l \frac{4 \pi}{2l+1} \frac{r^l}{r_i^{l+1}}
Y_{lm}^*({\hat r}_i) Y_{lm}({\hat r}) ,
\end{eqnarray}
for $i=2, \dots, N$. The Schr\"odinger equation Eq. (\ref{schrodinger})
may then be rewritten as
\begin{eqnarray}
[&-&\frac{1}{2} \bigtriangledown^2 -\frac{Z}{r}
+ r \sum_{m=-1}^{1}Y_{1m}({\hat r})
g_m({\bf X})] \Psi({\bf r}, {\bf X})  \nonumber \\
&+&H_{Z-1}^{N-1}({\bf X}) \Psi({\bf r}, {\bf X})
= E \Psi({\bf r}, {\bf X}) ,
\end{eqnarray}
where
\begin{eqnarray}
g_m({\bf X}) = \frac{4 \pi}{3} \sum_{i=2}^N \frac{1}{r_i^2}
Y_{1m}^*({\hat r}_i),
\end{eqnarray}
and $H_{Z-1}^{N-1}({\bf X})$ is the Hamiltonian for
$N-1$ electrons in
an external potential of $-(Z-1)/r$.
The first relation in Eq. (\ref{a-Psi(0)}) follows readily
by equating the coefficients of $r^{-1}$ terms,
and the second one by equating the coefficients
of $r^0 Y_{1m}({\hat r})$ terms.

The density is defined as
\begin{eqnarray}
\rho({\bf r})= N \int d{\bf X} |\Psi({\bf r}, {\bf X})|^2 ,
\end{eqnarray}
where $\int d{\bf X}$ denotes $\sum_s 
\int d{\bf x}_1 \dots d{\bf x}_N$.
Bear in mind that we are considering  
spherically symmetric or sphericalized systems.
Employing Eq. (\ref{wavefunction}) in conjunction with
Eq. (\ref{a-Psi(0)}) and the relation
\begin{eqnarray}  \label{Ylm}
\int Y^*_{1m}({\hat r}) Y_{1'm'}({\hat r}) d{\hat r} = \delta_{l l'}
\delta_{m m'} ,
\end{eqnarray} 
we obtain the density as 
\begin{eqnarray}
\rho(r)=& & N \int d{\bf X} 
\biggl ( |\Psi(0, {\bf X})|^2 (1-Zr)^2   \nonumber \\
&+&2 r^2 Re[\Psi^*(0, {\bf X}) b({\bf X})]   \nonumber \\
&+& r^2\sum_{m=-1}^{1}\frac{1}{4 \pi}|a_{1m} ({\bf X})|^2 \biggr )
+O(r^3) .
\end{eqnarray}
Evidently
\begin{eqnarray}
\rho(0) = N \int d{\bf X} |\Psi(0, {\bf X})|^2 .
\end{eqnarray}
One then can see that the density behaves as
\begin{eqnarray} \label{rho1}
\rho(r) &=& (1-Zr)^2  \rho(0)  
+ r^2 2N \int d {\bf X}
Re[\Psi^*(0, {\bf X}) b({\bf X})]    \nonumber \\
& &+ r^2 N \int d {\bf X}
\sum_{m=-1}^{1}\frac{1}{4 \pi}|a_{1m} ({\bf X})|^2 
+ O(r^3) .
\end{eqnarray}

Once again, with the assumption of smoothness, the orbital of the $S$
system may be written as
\begin{eqnarray} \label{Sorbital}
\phi_{nlm}({\bf r})= r^l [ && A_{nl} 
+ B_{nl} r + C_{nl} r^2   \nonumber \\
&& +D_{nl} r^3 + \dots ] Y_{lm} ({\hat r}).
\end{eqnarray}
Similarly the coefficients in the orbital
satisfy the cusp relations:
\begin{eqnarray} \label{B00-A00}
&&B_{n0} + Z A_{n0} =0 ,   \nonumber \\
&&2B_{n1} + Z A_{n1} =0 .
\end{eqnarray}
The density of the $S$ system,
which is supposed to be equivalent to that
of the interacting system, is related to the orbitals
as
\begin{eqnarray}
\rho ({\bf r})= \sum_{nlm} |\phi_{nlm}({\bf r})|^2.
\end{eqnarray}
In the
preceding summation over the one-particle states
and the remaining ones in the paper, 
the spin summation is implicitly included.
Substitution from Eq. (\ref{Sorbital}) leads to
\begin{eqnarray}
\rho (r)=&& \sum_{nlm} r^{2l} 
[|A_{nl}|^2  + 2Re(A^*_{nl}B_{nl}) r  \nonumber \\
&&+|B_{nl}|^2 r^2 + 2Re(A^*_{nl}C_{nl}) r^2 ] |Y_{lm}({\hat r})|^2.
\end{eqnarray}
Now employing Eq. (\ref{B00-A00}), we 
obtain
\begin{eqnarray} \label{rho4}
\rho (r) &=& \rho(0) (1-Zr)^2 +
r^2 \frac{1}{4 \pi} \biggl [ 2\sum_{n} Re (A_{n0}^*C_{n0})  \nonumber \\
& &+ 3\sum_{n}
|A_{n1}|^2 \biggr ] +O(r^3) .
\end{eqnarray}
The fact that $\rho (r)$ is a radial function, together
with Eq. (\ref{Ylm}), has also been employed to obtain Eq. (\ref{rho4}). 

\section{Electron-interaction field 
${\bf {\cal E}}_{ee} ({\bf r})$ near
the nucleus}

In this section, we show that the electron-interaction field 
${\bf {\cal E}}_{ee} ({\bf r})$ vanishes at the nucleus.
This can be understood on physical grounds. The pair-correlation
density, $g({\bf r}, {\bf r}')$, the quantal source of the 
field ${\cal E}_{ee}(r)$, is in general not spherically symmetric
about the electron position at ${\bf r}$. However, for an electron
at the nucleus, this charge distribution is spherically symmetric.
Hence, the field due to this charge as obtained by Coulomb's law,
must vanish at the nucleus. This result can also be proved
mathematically.

For spherically symmetric systems, 
the field ${\bf {\cal E}}_{ee} ({\bf r})$
can only have a radial component, and can only depend
upon the radial coordinate, i.e. ${\bf {\cal E}}_{ee} ({\bf r})
= {\hat r}{\cal E}_{ee, r}(r)$. Thus, by employing the identity
\begin{eqnarray} \label{id}
{\hat r} \cdot \bigtriangledown
=\partial/\partial r ,
\end{eqnarray}
one has, from Eq. (\ref{eee1}),
\begin{eqnarray}
{{\cal E}}_{ee, r}(r) = -
\int d{\bf r}' g({\bf r}, {\bf r}')  \frac{\partial}{\partial r}
\frac{1}{|{\bf r} - {\bf r}'|} .
\end{eqnarray} 
According to Eq. (\ref{wavefunction}), obviously $g({\bf r}, {\bf r}')$ 
is well behaved (not divergent) at ${\bf r} =0$. 
Therefore, 
\begin{eqnarray} \label{Eee=0}
{{\cal E}}_{ee, r}(r)|_{r=0}
= &&- \frac{4 \pi}{3} 
\int d{\bf r}' g(0, {\bf r}')  
\frac{1}{r'^2}    \nonumber \\
&& \sum_{m=-1}^{1} Y_{1m}({\hat r})
Y_{1m}^*({\hat r}') \nonumber \\
= && 0 .
\end{eqnarray}
(For open shell atoms, the central field approximation
means taking the spherical average of the field. Thus, in this
approximation, the field is again radial and depends only on
the radial coordinate.)

Now since ${\bf \cal E}_{ee}({\bf r})$ is the negative gradient of 
the scalar potential function $W_{ee}({\bf r})$, it 
is evident that there can be no linear term in this function. 
Thus, there is no contribution to the linear term of $v_{ee}(r)$
due to Pauli and Coulomb correlations. These correlations
contribute quadratically. As an example,
see Ref. \cite{slamet}.

Arguments similar to those above can also be made to prove
that the Pauli ${\bf \cal E}_{x}({\bf r})$ and Coulomb
${\bf \cal E}_{c}({\bf r})$ fields defined
as
\begin{eqnarray} \label{ex}
{\bf \cal E}_{x,c}({\bf r}) = \int d {\bf r}' 
\rho_{x,c}({\bf r}, {\bf r}')
\frac{{\bf r} - {\bf r}'}{|{\bf r} - {\bf r}'|^3} ,
\end{eqnarray}
vanish at the nucleus. (By the notation ${\bf \cal E}_{x,c}({\bf r})$
we mean the Pauli ${\bf \cal E}_{x}({\bf r})$ and 
Coulomb
${\bf \cal E}_{c}({\bf r})$ fields due to the Fermi $\rho_x({\bf r}, {\bf r}')$
and Coulomb $\rho_c({\bf r}, {\bf r}')$ holes, respectively.)
Once again, for example, for the electron
position at the nucleus, the Fermi and Coulomb holes are
spherically symmetric about the electron. Hence, the 
corresponding fields must vanish there. Hence, these
individual correlations contribute to the quadratic structure
of $v_{ee}(r)$ via the work $W_x(r)$ and $W_c(r)$ \cite{slamet}.

The classical Hartree field which is defined
as 
\begin{eqnarray} \label{eh}
{\bf \cal E}_{H}({\bf r}) = \int d {\bf r}' \rho({\bf r}')
\frac{{\bf r} - {\bf r}'}{|{\bf r} - {\bf r}'|^3} ,
\end{eqnarray}
corresponding to the Hartree potential,
$v_H({\bf r}) = \int d {\bf r}' \rho({\bf r}')/|{\bf r} - {\bf r}'|$,
also vanishes at the nucleus because the density $\rho({\bf r})$ is
spherically symmetric.

In conclusion, the electron-interaction 
field ${\bf \cal E}_{ee}({\bf r})$,
and its Hartree ${\bf \cal E}_{H}({\bf r})$, Pauli
${\bf \cal E}_{x}({\bf r})$, and Coulomb ${\bf \cal E}_{c}({\bf r})$
components all vanish at the nucleus.

\section{Correlation-kinetic field 
${\bf {\cal Z}}_{t_c} ({\bf r})$ near the nucleus}

We shall first present  
the calculation for the kinetic fields of the interacting system
and the {\em S} system separately in Subsec. A and Subsec. B.
In Subsec. C, we present that for the correlation-kinetic 
field ${\bf {\cal Z}}_{t_c} ({\bf r})$.

The summations over the dummy cartesian indices are assumed in this
section.

\subsection{Kinetic field ${\bf z}({\bf r})$ 
of the interacting system}

In order to make our presentation simpler,
we rewrite Eq. (\ref{wavefunction}) as
\begin{eqnarray}  \label{wavefunction2}
\Psi({\bf r}, {\bf X}) = && \Psi(0, {\bf X})(1-Zr)
+ {\bf \alpha}({\bf X}) \cdot {\bf r} ( 1 - Zr/2)  \nonumber \\
&& + \beta_{ij}({\bf X}) r_i r_j,
\end{eqnarray}
where we have made use of Eq. (\ref{a-Psi(0)}).
The relations between ${\bf \alpha}({\bf X})$ and $a_{1m}({\bf X})$,
and $\beta_{ij}({\bf X})$ and $b({\bf X})$, $b_{2m}({\bf X})$ are 
simply those between the cartesian tensors and spherical tensors.
By substituting Eq. (\ref{wavefunction2}) into Eq. (\ref{gamma}),
we have
the structure of the $\gamma({\bf r}, {\bf r}')$ 
for small $r$, $r'$ as
\begin{eqnarray}  \label{gammar=0}
\gamma({\bf r}, {\bf r}') &=& \rho(0)(1-Zr)(1-Zr') \nonumber \\ 
&&+(1-Zr)(1 -Zr'/2) {\bf A}^* \cdot {\bf r}'  \nonumber \\
&& +(1-Zr')(1 -Zr/2) {\bf A}\cdot {\bf r}   \nonumber \\
&& + E_{ij}^*(1-Zr/2)(1-Zr'/2)r_i r_j'  \nonumber \\
&& +[D_{ij}^*(1-Zr) +F_{lij}^*r_l] r_i' r_j'  \nonumber \\
&& +[D_{ij}(1-Zr') +F_{lij} r_l'] r_i r_j ,
\end{eqnarray}
where
\begin{eqnarray}
A_i= N \int d{\bf X} \Psi(0, {\bf X})  \alpha_i^*({\bf X}),
\end{eqnarray}
\begin{eqnarray}
E_{ij} = N \int d{\bf X} \alpha_i({\bf X}) \alpha_j^*({\bf X}),
\end{eqnarray}
\begin{eqnarray}
D_{ij}= N \int d{\bf X} \Psi(0, {\bf X}) \beta_{ij}^*({\bf X}),
\end{eqnarray}
\begin{eqnarray}
F_{ijl} = N \int d{\bf X} \alpha_i({\bf X}) \beta_{jl}^*({\bf X}).
\end{eqnarray}
It is necessary to mention that the preceding expression for
$\gamma({\bf r}, {\bf r}')$ is not limited to
the spherically symmetric or sphericalized systems, but valid in general.
For systems with central-reflection symmetry, Eq. (\ref{gammar=0})
reduces to
\begin{eqnarray}  \label{gammar=0again}
\gamma({\bf r}, {\bf r}') &=& \rho(0)(1-Zr)(1-Zr') \nonumber \\
&+& E_{ij}^*(1-Zr/2)(1-Zr'/2)r_i r_j'  \nonumber \\
&+& D_{ij}^*(1-Zr)r_i' r_j' + D_{ij}(1-Zr')r_i r_j .
\end{eqnarray}
We note that 
the second, third and fourth terms on the right hand side of the 
above expression are absent in the literature \cite{bingel,bader}. 
Also for completeness, we give the following relations:
\begin{eqnarray}
D_{ii} = && 3N \int d{\bf X} \Psi^*(0, {\bf X}) b({\bf X}) ,  \nonumber \\
E_{ii} = && 3N \int d{\bf X} \frac{1}{4 \pi} |a_{1m}({\bf X})|^2 .
\end{eqnarray}
Correspondingly, the expression of Eq. (\ref{rho1}) for the density can
be rewritten as
\begin{eqnarray} \label{density}
\rho(r) = \rho(0)(1-Zr)^2 + \frac{1}{3}(E_{ii} + 2 D_{ii}) r^2 + O(r^3).
\end{eqnarray}

The derivatives in calculating the kinetic energy density tensor
via Eq. (\ref{tij0})
are tedious but straightforward. After performing them, we arrive at
\begin{eqnarray}  \label{tij}
&& t_{ij}({\bf r}) = \frac{1}{2}Re \biggl [ Z^2(\rho(0)
+{\bf A} \cdot {\bf r} )\frac{r_i r_j}{r^2}   \nonumber \\
&&- \frac{Z}{2r}(2-Zr)(r_iA_j + 
r_j A_i)    
+ E_{ij} (1-Zr)   \nonumber \\
&&-\frac{Z r_l}{2r} [ r_i (E_{lj} + 4D_{lj}) + r_j (E_{li}
+4D_{li} ) ]   \nonumber \\
&& +2(F_{ijl} +F_{jil}) r_l \biggr ]
+ O(r_i^2) .
\end{eqnarray}
The corresponding kinetic energy density is
\begin{eqnarray}  \label{t(r)}
{t}(r) &=& {t}_{ii} (r)  
=\frac{1}{2} [ Z^2 \rho(0) +E_{ii}]  \nonumber \\
&& -\frac{2Z}{3} r [ D_{ii}+E_{ii}]
+ O(r^2) .
\end{eqnarray}
The correctness of the expression for $t(r)$ is independently
confirmed via its definition as
\begin{eqnarray}  \label{t-definition}
t({\bf r}) =
\frac{1}{2} N \int d {\bf X} \bigtriangledown \Psi^*({\bf r}, {\bf X})
\cdot \bigtriangledown \Psi({\bf r}, {\bf X}).
\end{eqnarray}

With the result for $t_{ij}({\bf r})$ given in Eq. (\ref{tij}),
one can obtain from Eq. (\ref{z-defin})
the kinetic field ${\bf z}({\bf r})$ to be
\begin{eqnarray}
{\bf z}({\bf r}) = 2Z^2 \rho(0)\frac{\bf r}{r^2} -
\frac{4}{3}Z(3D_{ii} +E_{ii})\frac{\bf r}{r} +O(r) ,
\end{eqnarray}
or,
\begin{eqnarray}  \label{z}
{\bf z}({\bf r}) =&& 2Z^2 
\rho(0)\frac{\bf r}{r^2}(1 + 2Zr/3)   \nonumber \\
&& -\frac{4}{3}Z[2t(0) +3D_{ii}] \frac{\bf r}{r} + O(r) ,
\end{eqnarray}
where we have made use of the following relation
derived from Eq. (\ref{t(r)}):
\begin{eqnarray}  \label{t(0)}
E_{ii} = 2t(0) - Z^2 \rho(0).
\end{eqnarray}

Equation (\ref{t(0)}) can also be employed to rewrite the expression
for the density in Eq. (\ref{density}) as
\begin{eqnarray} \label{rho3}
{\rho}(r) &=&
\rho(0) \biggl \{ 1-2Zr 
+ \frac{2}{3\rho(0)}[Z^2 \rho(0) + t(0)] r^2 \biggr \}  \nonumber \\
& +& \frac{2}{3} r^2 D_{ii} +O(r^3) .
\end{eqnarray}
This expression for the density is employed later.

\subsection{Kinetic field ${\bf z}_s({\bf r})$ of the {\em S} system}

The kinetic energy density tensor for the {\em S} system is
expressed in terms of the orbitals as
\begin{eqnarray}
t_{s,ij}({\bf r}) = \frac{1}{2} Re \sum_{nlm}
\biggl [\frac{\partial}{\partial r_i} 
\phi_{nlm}^*({\bf r}) \biggr ]
\biggl [\frac{\partial}{\partial r_j} \phi_{nlm}({\bf r}) \biggr ] .
\end{eqnarray} 
Substituting Eq. (\ref{Sorbital}) into the above expression
and making use of Eq. (\ref{B00-A00}), 
one has,
\begin{eqnarray}  \label{tsij}
t_{s,ij}({\bf r})   &&= \frac{1}{2} 
Z^2 \rho(0) r_i r_j /r^2  \nonumber \\
&& +\frac{3}{8 \pi}\sum_n |A_{n1}|^2( \delta_{ij} 
-Zr \delta_{ij} - Zr_ir_j/r)  \nonumber \\
&& -2Z \sum_n \frac{1}{4 \pi}
Re(A_{n0}^*C_{n0})r_ir_j/r + O(r_i^2).
\end{eqnarray}
The kinetic energy density for the {\em S} system correspondingly is
\begin{eqnarray} \label{ts(r)}
t_s(r) = && t_{s,ii} ({\bf r})  \nonumber \\
= && \frac{1}{2}Z^2 \rho(0) 
- 2Z r \sum_n \frac{1}{4 \pi} Re A_{n0}^*C_{n0}   \nonumber \\
&& + \frac{1}{2}(3-4Zr) 
\sum_{n} \frac{3}{4 \pi} |A_{n1}|^2 
+O(r^2) .
\end{eqnarray}
This expression for $t_s(r)$ is independently confirmed via its
definition
\begin{eqnarray}  \label{ts-definition}
t_s({\bf r})= \frac{1}{2} \sum_{nlm} 
\bigtriangledown \phi_{nlm}^*({\bf r})
\cdot \bigtriangledown \phi_{nlm}({\bf r}).
\end{eqnarray}
(We note that the leading term $t_s(0)$ obtained from Eq. (\ref{ts(r)})
has been obtained previously \cite{janosfalvi}.) 

The calculation for the kinetic field $z_{s,i} ({\bf r})=
2 \frac{\partial}{\partial r_j} t_{s,ij}({\bf r})$ is now straightforward,
which yields
\begin{eqnarray}
{\bf z}_{s}({\bf r}) = && 2Z^2 \rho(0) \frac{\bf r}{r^2}
- 12Z \sum_{n} \frac{1}{4 \pi} Re (A_{n0}^*C_{n0}) 
\frac{\bf r}{r}  \nonumber \\
&& - 4Z \sum_{n} \frac{3}{4 \pi} 
|A_{n1}|^2 \frac{\bf r}{r} + O(r),
\end{eqnarray}
or, by the use of Eq. (\ref{ts(r)}),
\begin{eqnarray} \label{zs}
{\bf z}_s({\bf r}) = && 2Z^2 \rho(0)\frac{\bf r}{r^2} (1+ 2Zr/3)
-4Z \frac{\bf r}{r}  \biggl [ 
\frac{2}{3} t_s(0)   \nonumber \\
&&+3 \sum_n \frac{1}{4 \pi}  
Re (A_{n0}^*C_{n0}) \biggr ] +O(r) .
\end{eqnarray}
Also by the use of Eq. (\ref{ts(r)}),
the expression of the density for the $S$ system in Eq. (\ref{rho4})
can be rewritten 
as
\begin{eqnarray} \label{rho5}
{\rho}(r) &=& {\rho}(0) \biggl \{
1-2Zr + \frac{2}{3 \rho(0)}[Z^2 \rho(0) + t_s(0)] r^2 \biggr \}  \nonumber \\
&+& 2r^2 \sum_{n} \frac{1}{4 \pi} Re (A_{n0}^*C_{n0})
+O(r^3).
\end{eqnarray}
This expression for the density is employed below.

\subsection{Correlation-kinetic field ${\bf {\cal Z}}_{t_c}({\bf r})$}

We now substitute 
Eq. (\ref{z}) and Eq. (\ref{zs}) into Eq. (\ref{Ztc1}), and arrive
at
\begin{eqnarray}  \label{Ztc2}
{\bf {\cal Z}}_{t_c}({\bf r}) 
&=& \frac{4Z}{\rho(r)} \frac{\bf r}{r}
\biggl [ \frac{2}{3} (t(0) -t_s(0)) +D_{ii} \nonumber \\
&&-3 \sum_{n} \frac{1}{4 \pi} Re (A_{n0}^*C_{n0}) \biggr ] +O(r).
\end{eqnarray}
Now the density of the $S$ system is  
equal to that of the interacting
system. Comparing Eq. (\ref{rho3}) and Eq. (\ref{rho5}) leads to
\begin{eqnarray}  \label{relation}
\frac{1}{2 \pi} \sum_{n} Re (A_{n0}^*C_{n0})
= \frac{2}{3} [t(0) - t_s(0) + D_{ii} ].
\end{eqnarray}
Making use of the above relation,
one obtains, from Eq. (\ref{Ztc2}),
\begin{eqnarray} \label{Ztc3}
{\bf {\cal Z}}_{t_c} ({\bf r}) = \frac{4Z}{3} \frac{1}{\rho(0)} 
[t_s(0) - t(0)] \frac{\bf r}{r}
+O(r) .
\end{eqnarray} 
In contrast to the electron interaction 
field ${\cal E}_{ee}({\bf r})$, the correlation-kinetic
field ${\cal Z}_{t_c} ({\bf r})$ is nonzero at the nucleus.
For an example, see Ref. \cite{topic}.
In passing, we mention Ref. \cite{liu} for a discussion of an approximate
relation of the property $\int d{\bf r} [t_s({\bf r}) - t({\bf r})]$
to $\rho(0)$.

\section{Behavior of $v_{ee}(r)$ near the nucleus}

Combining the results in Eqs. (\ref{Eee=0}) and (\ref{Ztc3}), one finally has,
from Eq. (\ref{vee-eee-ztc}),
\begin{eqnarray}   \label{vee(r)2}
\bigtriangledown v_{ee}(r) = \frac{4Z}{3 \rho(0)}  [t(0) - t_s(0)]
\frac{\bf r}{r}+O(r),
\end{eqnarray} 
which also indicates the result shown in Eqs. (\ref{vee}) and (\ref{beta}).
The classical Hartree potential due to the density,
$v_H({\bf r})$,
is usually separated out from $v_{ee}({\bf r})$
and the remaining part is referred as the HKS-DFT exchange-correlation 
potential $v_{xc} ({\bf r})$.
Since it is evident from the structure of the Hartree 
field ${\bf \cal E}_H({\bf r})$ that there is no linear term 
in the Hartree potential $v_H(r)$,
Eqs. (\ref{vee}) and (\ref{beta}) imply that
\begin{eqnarray} \label{vxc(r)}
v_{xc}(r)= v_{xc}(0) +
\frac{4Z}{3\rho(0)} [t(0) -t_s(0)] r + O(r^2).
\end{eqnarray}

Our calculations hence show that the 
linear term of $v_{ee}(r)$ (or equivalently $v_{xc}(r)$) 
arises {\em solely} from
the correlation-kinetic effects.

\section{Behavior of the HKS exchange $v_x({\bf r})$ and correlation
$v_c({\bf r})$ potentials near the nucleus}

In HKS-DFT, the exchange-correlation potential $v_{xc}({\bf r})$
is usually split into an exchange $v_{x}({\bf r})$ and a
correlation $v_{c}({\bf r})$ component. These potentials have
been provided a rigorous interpretation in terms
of the Pauli, Coulomb, and correlation-kinetic
fields via Q-DFT \cite{sahni,levyqian}
as follows
\begin{eqnarray}\label{vx}
v_x({\bf r}) &=& - \int_\infty^{\bf r} [{\bf {\cal E}}_x ({\bf r}')
-{\bf Z}_{{t_c},1} ({\bf r}') ] \cdot d{\bf l}' , \\
v_c({\bf r}) &=& - \int_\infty^{\bf r} [{\bf {\cal E}}_c ({\bf r}')
-{\bf Z}_{{t_c},2} ({\bf r}') -{\bf Z}_{{t_c},3} ({\bf r}') - \dots ]
\cdot d{\bf l}',   \nonumber \\
\end{eqnarray}
where ${\bf Z}_{{t_c},1} ({\bf r})$, and ${\bf Z}_{{t_c},2} ({\bf r})$, etc.,
are the lowest order and higher order components,
respectively, of the correlation-kinetic
field ${\bf Z}_{t_c} ({\bf r})$ as obtained via the adiabatic coupling
constant perturbation scheme \cite{sahni,levyqian}.
Thus, we see that HKS exchange potential $v_{x}({\bf r})$ is
representative not only of Pauli correlation but
also of the lowest-order correlation-kinetic effects. The
correlation potential $v_{c}({\bf r})$ is representative of
Coulomb correlations and higher-order correlation-kinetic effects.

At this stage, it also becomes clear that the entire derivation
of this paper can be reformulated within adiabatic coupling
constant perturbation scheme \cite{harris,levyqian}. 
In such a scheme one has, instead of the result in
Eq. (\ref{Ztc2}) for ${\bf {\cal Z}}_{t_c}({\bf r})$,
\begin{eqnarray} \label{Ztclamda}
{\bf {\cal Z}}_{t_c}^\lambda ({\bf r}) = \frac{4Z}{3 \rho(0)}
[t_s(0) - t^\lambda (0)] \frac{\bf r}{r}
+O(r) ,
\end{eqnarray}
where $\lambda$ is the coupling constant. Define
$t_x({\bf r})$ as the first order correction
to $t_s({\bf r})$ in the adiabatic coupling constant perturbation
scheme,
and $t_c({\bf r})=  [t({\bf r}) -t_s({\bf r})]- t_x({\bf r})$,
we have from Eq. (\ref{Ztclamda})
\begin{eqnarray} \label{ZTC1}
{\bf {\cal Z}}_{t_c,1} ({\bf r}) &&= \frac{4Z}{3 \rho(0)} t_x(0)
\frac{\bf r}{r}
+O(r) ,    \\
{\bf {\cal Z}}_{t_c,2} ({\bf r}) +&& {\bf {\cal Z}}_{t_c,3} ({\bf r})
+ \dots = \frac{4Z}{3 \rho(0)} t_c(0)
\frac{\bf r}{r}
+O(r) .
\end{eqnarray}
Thus the work done in these fields is
\begin{eqnarray}  \label{WTC1}
W_{t_c,1}(r)= W_{t_c,1}(0) +
\frac{4Z}{3 \rho(0)} t_{x}(0)r + O(r^2),
\end{eqnarray}
and
\begin{eqnarray}  \label{WTC23}
W_{t_c,2}(r)+ W_{t_c,3}(r) &&+ \dots = W_{t_c,2}(0)
+ W_{t_c,3}(0) + \dots   \nonumber \\
&&+
\frac{4Z}{3 \rho(0)}Z t_{c}(0)r + O(r^2).
\end{eqnarray}
Since we have shown in Sec. IV that  ${\bf {\cal E}}_x (r)$
and ${\bf {\cal E}}_c (r)$ both vanish at the nucleus, 
we obtain the expression for $v_x(r)$ and $v_c(r)$ near
the nucleus as
\begin{eqnarray}\label{vx,c}
v_{x,c}(r)= v_{x,c}(0) +
\frac{4Z}{3 \rho(0)} t_{x,c}(0)r + O(r^2).
\end{eqnarray}
Thus, the $v_x({\bf r})$ approaches the nucleus linearly, the term
arising solely due to lowest-order correlation-kinetic
effects. The $v_c({\bf r})$ also approaches the nucleus linearly,
the contribution to this term arising from higher-order 
correlation-kinetic effects.

Finally, it is worth mentioning that, as is well known, there
is another definition of the kinetic energy density of
the interacting and noninteracting systems \cite{cohen}:
\begin{eqnarray}  \label{t-tilde-definition}
{\tilde t}({\bf r}) =&&
-\frac{1}{2} N \int d {\bf X} \Psi^*({\bf r}, {\bf X})
\bigtriangledown^2 \Psi({\bf r}, {\bf X}) ,  \nonumber \\
{\tilde t}_s({\bf r})=&& -\frac{1}{2} \sum_{nlm}
\phi_{nlm}^*({\bf r})
\bigtriangledown^2 \phi_{nlm}({\bf r}).
\end{eqnarray}
Though ${\tilde t}({\bf r})$ and ${\tilde t}_s({\bf r})$ yield respectively
the same total kinetic energies for finite systems as
$t({\bf r})$ and $t_s({\bf r})$ defined in
Eqs. (\ref{t-definition}) and (\ref{ts-definition}), they differ from
$t({\bf r})$ and $t_s({\bf r})$:
\begin{eqnarray}
t({\bf r})= && {\tilde t}({\bf r})
+ \frac{1}{4} \bigtriangledown^2 \rho({\bf r}) ,  \nonumber \\
t_s({\bf r})= && {\tilde t}_s({\bf r})
+ \frac{1}{4} \bigtriangledown^2 \rho({\bf r}).
\end{eqnarray}
The second term on the right hand side
of the preceding relations, 
which are singular at the nucleus are the same for
the interacting and non-interacting 
systems. Therefore, all the results in this paper
hold equivalently in the alternative definition. For instance,
one also has, in parallel to Eq. (\ref{vxc(r)}),
\begin{eqnarray}
v_{xc}(r)= v_{xc}(0) +
\frac{4Z}{3\rho(0)} [{\tilde t}(0) -{\tilde t}_s(0)] r + O(r^2).
\end{eqnarray}

\section{Conclusions}

In conclusion we note (i) the electron-interaction potential 
$v_{ee}({\bf r})$ of the model electrons is finite at the nucleus. 
Correlations due to the Pauli exclusion principle, Coulomb repulsion, and
correlation-kinetic effects {\em all} contribute to this
value. In this paper we have proved the following via Q-DFT: (ii)
that the function $v_{ee}({\bf r})$ approaches the nucleus linearly;
(iii) there are no contributions to this linear term due to
Pauli and Coulomb correlations; (iv) the Pauli and Coulomb correlations,
however, contribute to the quadratic term of this function; (v) the
linear term in the structure of $v_{ee}({\bf r})$ is solely due to
correlation-kinetic effects. This contribution is determined analytically.
We have also obtained by application of adiabatic coupling constant
perturbation theory via Q-DFT (vi) the near nucleus structure
of the HKS-DFT exchange $v_{x}({\bf r})$ and correlation
$v_{c}({\bf r})$ potentials. These potentials also approach
the nucleus linearly: the linear term of 
$v_{x}({\bf r})$ is solely due to the lowest order correlation kinetic
effects; the linear term of $v_{c}({\bf r})$ is solely 
due to higher-order correlation-kinetic contributions.
We further note that the results in this work are valid in general whether
the system is an atom, molecule or a solid.

In previous work \cite{sham}, it has been shown that in the classically
forbidden region of spherically symmetric systems, the contribution to
the asymptotic structure of $v_{ee} ({\bf r})$ due to
Pauli correlations is of $O(1/r)$, that of Coulomb correlations
of $O(1/r^4)$, and those of correlation-kinetic effects
of $O(1/r^5)$. Thus, together with the present results, the
structure of $v_{ee} ({\bf r})$ in the asymptotic
near nucleus and classically forbidden regions of such systems is
now known analytically. (For the analytic asymptotic
structure of $v_{ee} ({\bf r})$ in the classically forbidden region at 
the semi-infinite metal-vacuum interface, and the contribution of each type
of correlation to this structure, we 
refer the reader to Ref. \cite{sahni,qiansahni3}.)

Finally, the result derived also has  
bearing on the construction
of approximate HKS-DFT electron-interaction 
energy functionals of the density.
In numerical codes employing such approximate functionals, the structure
of the corresponding $v_{ee} ({\bf r})$ is obtained as the functional
derivative. The derivative, being dependent on the density,
exhibits a sharp cusp \cite{fig} at the nucleus just as does the density. 
However, calculations \cite{umrigar,slamet} based on accurate wavefunctions
show the slope of the function $v_{ee} ({\bf r})$ at and near the
nucleus to be extremely small.
Thus,
these approximate energy functionals ought now be modified so that
their functional derivative has the correct near nucleus asymptotic structure.

We conclude by noting that the results of the paper can
also be derived via the HKS-DFT \cite{qian}. However, the advantage
of working within the framework of Q-DFT is that the near nucleus structure
of $v_{ee} ({\bf r})$ can be understood as a function of the separate
electron correlations that contribute to it.

ZQ acknowledges support from the Chinese National Science Foundation
under Grant No. 10474001. The work of VS was supported in part by
the Research Foundation of CUNY.

\end{document}